\documentclass{article}

\usepackage[margin=1in]{geometry}
\usepackage{xcolor}
\usepackage{colortbl}
\usepackage{graphicx}
\usepackage{wrapfig}
\usepackage{comment}
\usepackage{multirow}
\usepackage{hhline}

\usepackage[breaklinks]{hyperref}                                                
\hypersetup{pdfborder = 0 0 0}                                                   
\hypersetup{pdftitle={Towards Exascale Scientific Metadata Management}}
\hypersetup{pdfauthor={Spyros Blanas and Suren Byna}}

\title{Towards Exascale Scientific Metadata Management}

\author{
Spyros Blanas \hspace{3cm} Surendra Byna\\
\small{\hspace{ 0.8cm} The Ohio State University \hspace{1cm} Lawrence Berkeley National Laboratory}\\
\small{\hspace{-0.5cm} blanas.2@osu.edu \hspace{3.4cm} sbyna@lbl.gov}
}
\date{}

\begin{document}

\maketitle

{
\abstract

\vspace{1em}
\noindent
Advances in technology and computing hardware are enabling scientists from all
areas of science to produce massive amounts of data using large-scale
simulations or observational facilities. In this era of data deluge, effective
coordination between the data production and the analysis phases hinges on the
availability of metadata that
describe the scientific datasets. 
Existing workflow engines have been capturing a
limited form of metadata to provide provenance information about the identity and
lineage of the data.
However, much of the data produced by simulations, experiments, and analyses still need to
be annotated manually in an ad hoc manner by domain scientists. 
Systematic and transparent 
acquisition of 
rich metadata
becomes a crucial prerequisite to sustain and accelerate the pace
of scientific innovation. 
Yet, ubiquitous and domain-agnostic metadata
management infrastructure that can meet the demands of extreme-scale
science is notable by its absence. 

\vspace{1em}
\noindent
To address this gap in scientific data management research and practice, we present
our vision for an integrated approach that (1) automatically captures and
manipulates in\-for\-ma\-tion-rich metadata while the data is being produced or
analyzed and
(2) stores metadata within each dataset to permeate metadata-oblivious
processes and to query metadata through established and standardized data
access interfaces.
We motivate the need for the proposed
integrated approach using applications from plasma physics,
climate modeling and neuroscience, and then
discuss research challenges and possible solutions.

}


\section{Introduction}

The observation or simulation of natural phenomena produces massive
datasets that are cumbersome to manage. 
Although many efforts are underway to address the research challenges of
storing and analyzing large scientific datasets, the research community
has paid less attention in using domain-agnostic metadata to improve
performance and scientific productivity.
Metadata is essential 
to automate scientific analysis tasks and workflows, and can bring  
logical data independence to scientific applications through
metadata-aware scientific management tools \cite{GrayOnSDM}. 
Information-rich metadata, such as low-resolution snapshots of datasets
and of results from historical analyses, have the potential to guide
scientists or runtimes to perform data analysis more efficiently. 
In addition, 
effective collaboration between teams frequently
hinges on succinctly conveying the salient properties of a large
dataset as metadata.
Therefore, systematic management of information-rich metadata can be a
catalyst for scientific productivity when progress depends on the
coordination of hundreds of scientific teams that span discipline
boundaries.
For instance, understanding of the effects of global warming on future
generations involves more than 1,200 scientists in the IPCC's AR-5
Working Group II alone \cite{AR-5_number}.
Unfortunately, the vision of systematic scientific metadata
management across heterogeneous analysis tools and science disciplines
largely remains unrealized. 

Conventional practice for scientific metadata management varies between
scientific domains, and sometimes even between individual applications.
The scientific metadata that are collected today can be classified in
three broad categories.
\emph{Descriptive metadata} are integrated within datasets and are largely
tailored to help scientists locate and read data variables. 
Even file format libraries and systems that are designed for
large-scale scientific data analysis such as ADIOS/BP \cite{adios} and
SciDB \cite{SciDB} store little metadata beyond quantile information for
each block of data.
Metadata conveyed in \emph{ad-hoc text annotations} have high
information density but are time-consuming to generate. 
Because of this productivity cost, datasets are selectively annotated
with ad-hoc information during data curation.
An unfortunate consequence is that many datasets that scientists
interact with daily are in intermediate, non-curated forms without any
metadata annotations.
\emph{Data provenance} metadata is automatically captured by workflow
systems and reveals information about the data generating process, any
data transformations, the historical analyses and their associated data
movement. 
However, data movement that is oblivious to the workflow system (such as
a file transfer) severs the connection between a dataset and its
provenance metadata because of their physical separation in different
storage locations.

\begin{figure*}
\centering
\includegraphics[width=0.9\textwidth]{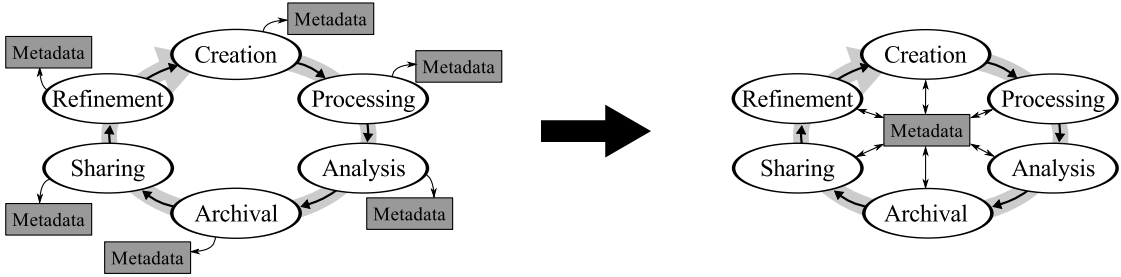}
\caption{Our vision for exascale scientific metadata management
integrates metadata information currently in disparate containers into a
single metadata acquisition and storage framework that integrates
with established scientific file formats.}
\label{fig:mdcycle}
\end{figure*}

Existing metadata management practices prove insufficient when
analyzing massive datasets using tens of thousands of CPU cores 
at leadership computing facilities.
At this compute scale and data volume, scientists require guidance on
how to perform data triage and decide what dataset fragments need
to be processed first or ignored.
Data analysis runtimes rely on external information from an
infrastructure expert to tune parallel I/O and optimally co-locate
dataset fragments that are analyzed simultaneously.
Data curators have to manually discover what metadata are available for
each dataset and data variable to verify whether it is accurate or
outdated.
We posit that scientific metadata ought to be captured automatically,
stored within the scientific dataset, and be accessed as frictionlessly
as regular data to quicken the pace of scientific discovery.
Our investigation is motivated by two questions:
\begin{itemize}
\item
What new metadata types can be captured systematically that will
accelerate scientific discovery?
\item
How can this metadata be acquired, stored, and queried efficiently?
\end{itemize}

In this paper, we categorize existing metadata management practices,
discuss use cases for information-rich metadata for accelerating
representative scientific applications
and present some research challenges in realizing this vision.
We elaborate on the current practices for managing metadata and discuss the
potential of integrated metadata management to impact the data-driven
discovery process in Section~\ref{sec:framework}.
We motivate the need for automated metadata acquisition
and management using scientific applications from 
plasma physics, climate modeling and analysis, and neuroscience in Section
\ref{sec:app-drivers}.
We discuss research challenges and possible solutions in acquiring, 
storing and accessing rich metadata in Section \ref{sec:approach}. We
discuss related work in Section \ref{sec:relwork} and conclude the paper
in Section \ref{sec:conclusions}.


\section{A vision for integrated metadata management}
\label{sec:framework}

\subsection{Metadata and data-driven discovery}

Data-driven scientific discovery often follows a cyclical pattern of
data creation, processing, analysis, archival, and sharing \cite{UKDA}. 
The shared dataset then becomes the seed for follow-up investigations
that further refine the research question or investigate new hypotheses.
This triggers the start of another iteration of the cycle.
Although metadata are generated and consumed in every step of this
cycle, the management of this metadata information across different
phases is unprincipled (shown on the left in Figure \ref{fig:mdcycle}).
A metadata management framework for extreme-scale science 
needs to capture metadata during all phases of the data life cycle
and expose that information to data generation or analysis tasks. 
This is shown on the right in Figure \ref{fig:mdcycle}.

We envision an integrated metadata management framework that augments 
scientific data with information that 
(1) is semantically richer, 
(2) is stored within each dataset and can be accessed via established
data access and querying interfaces, and
(3) is acquired, propagated, and managed automatically.
Richer metadata become a necessity
because the cost of sifting through an entire dataset to extract
a particular property of the data grows exponentially compared to
the cost of storing and transferring this property as metadata.
By storing metadata within a dataset, metadata can seamlessly
propagate through file-centric analyses and be easily transferred to
other scientists or computing facilities.
By relying on established data access interfaces for metadata access and
storage, next-generation scientific metadata management is unfettered from
particular analysis tools or workflow engines.

We identify four critical roles for information-rich metadata in the
data-driven scientific discovery cycle: Metadata can be used to appraise
a new dataset or acquaint a scientist with a curated dataset, develop,
and deploy a meaningful analysis, acquire insights and propagate them
across analysis, and accelerate data curation. We now elaborate on these
four roles.

\subsubsection*{Appraise and acquaint with a dataset}
In the first phase of the cycle, a scientist is given access to an
interesting dataset and uses metadata to acquaint himself with the
dataset and appraise its scientific value. 
Currently, datasets embed a limited form of metadata, 
such as a list of the data objects that the dataset
consists of, schema information, and the shape and dimensionality of
each data object. 
This metadata information is commonly insufficient, and 
additional insights are derived by computing simple statistics on
particular data objects and/or selectively visualizing data fragments of
interest.
In addition, scientists communicate with the data producer(s) for
additional information about the dataset. Collectively, the information
contained in these summarizations, visualizations, e-mails
and meeting notes semantically annotates the dataset.

Exascale science requires richer metadata that is embedded within
scientific datasets.
An information-rich metadata management framework should allow scientists to
understand a dataset through multi-resolution summaries that have been
retained from prior analyses and are embedded in the data.
Metadata-aware analysis tools can leverage multi-resolution summaries 
and accelerate exploratory data analysis.
The summaries of the datasets from prior analyses
can be visualized instantly. This substantially increases
the scientific re-use value of the dataset.

\subsubsection*{Develop and deploy analyses}
After the salient properties of the dataset are extracted, the scientist
decides on the appropriate domain-specific analysis that will be
performed and starts collecting the tools or developing the code to
perform the analysis with tens of thousands of CPU cores. 
Scientists need to debug and verify the output of an analysis using
a small representative sample of the dataset.
In this process, the scientist manually specifies additional metadata
that is specific to the dataset (such as what is a meaningful sample for
verification) and to the analysis infrastructure (such as particular file
locations or the available memory per node).
Larger and larger dataset samples are analyzed and verified, until there
is confidence that this analysis is correct and well-tuned for a
production run.

The proposed integrated scientific metadata framework shall retain
information on what analyses have been previously evaluated on a
dataset. 
The framework shall store qualitative information that is specific to
each analysis, such as dataset access patterns and access correlations
across datasets.
In addition, the framework has to capture the observed I/O and CPU
performance, as well as hardware configuration details, such as number
of CPU cores used, memory, parallel file system configuration, computer
topology, etc.
Richer metadata information about an analysis and its execution
environment allows scientists to spend less time debugging and tuning 
large-scale data processing pipelines. During deployment, scientific
tools can use the performance data stored as metadata to further
optimize and refine how an analysis will be executed.

\subsubsection*{Acquire and propagate insights}
If the analysis is orchestrated using a workflow system, provenance
information is automatically captured in a separate data store, which
may be a relational database management system. Existing systems such as 
MPO (Metadata, Provenance, and Ontology) \cite{MPO} is an example of this
approach. The representation and
storage of provenance information is specific to the workflow system
used by the scientist.
Otherwise, the scientist must manually acquire, manage and propagate
metadata information using scripts to a metadata-aware analysis task.
Tasks access metadata information via a proprietary interface that is
specific to the storage layout and representation of the metadata.

Our vision is to embed sophisticated metadata acquisition and management
primitives within popular scientific file formats. Thus, current applications
will automatically capture information-rich metadata, such as cross-dataset
correlations, without modifying existing analysis tools.
Scientific metadata are stored within the dataset and are managed by the
file format library. Metadata can then be accessed through 
the same established access interfaces for regular data objects.

\subsubsection*{Curate data}
The curation process aims to preserve selected datasets and 
their metadata for reuse.
Curation is commonly performed manually and selectively by the domain
expert.
The first step is to discard data that will not be curated, and
transform the dataset to a format that is appropriate for long-term
storage.
This metadata-oblivious transformation inadvertently strips
datasets from all but the most essential metadata (such as variable type
information). 
Metadata information that is stored separately, such as
lineage information from a workflow system, becomes stale.
Currently, the last step of the curation process augments the curated
dataset with semantically richer metadata information that the domain
expert deems necessary for future analyses. Such metadata are
commonly curated as ad-hoc annotations to the dataset.

With our vision of embedding metadata within the dataset, metadata curation becomes an
integral and concurrent part of the data curation process. 
Automatically acquiring, managing, and propagating metadata within
existing analysis pipelines will produce metadata-rich datasets with
minimal involvement from domain experts, 
which can thus relieve the need for
extensive ad-hoc metadata annotations during data curation. 
We anticipate that transparent metadata acquisition and
management will make scientists more likely to consume,
produce, and share metadata-rich datasets.
This expectation is supported by anecdotal evidence that 
scientists become more reluctant to share datasets as the metadata
annotation requirements become more onerous \cite{CornillonGS03}.

Scientists can use the curated access metadata to reconstruct an
audit trail of accesses to the dataset and understand what portions of
the data other scientists have analyzed extensively, and what analyses
have been performed.
Infrastructure providers of supercomputing systems can use performance-related
aspects of curated metadata to make informed procurement and deployment
decisions.
The broader scientific community and the general public can use 
information about dataset and fragment popularity to identify
influential datasets, data providers and projects, and acknowledge their
role in advancing large-scale data-driven science.

\begin{table*}
\newcommand{\mc}[3]{\multicolumn{#1}{#2}{#3}}
\newcommand{\oc}[1]{\multicolumn{1}{c|}{#1}}
\centering
\begin{tabular}{|p{0.1in}|c|c|c|}
\cline{3-4}
  \mc{2}{c|}{ } & \mc{2}{c|}{\multirow{2}{20em}[0em]{\centering \textbf{Metadata storage}}}
\\
  \mc{2}{c|}{ } & \mc{2}{c|}{ }
\\
\cline{3-4}
  \mc{2}{c|}{ } 
  & \multirow{2}{6em}[0em]{\centering \textbf{Segregated}}
  & \multirow{2}{6em}[0em]{\centering \textbf{Integrated}}
\\
  \mc{2}{c|}{ } 
  & 
  & 
\\
\hline
 \multirow{2}{0em}[-0.5em]{\rotatebox[origin=c]{90}{\textbf{Metadata source}}}
 & \multirow{3}{12em}[0em]{\centering \textbf{Human-generated}}
 & \multirow{3}{12em}[0em]{\centering Ad-hoc annotations}
 & List of data objects
\\
 &
 &
 & Array dimensionality and shape
\\
 &
 &
 & Schema and chunking information
\\
\hhline{~---}
 & \multirow{5}{12em}[0em]{\centering \textbf{Automatically acquired}}
 & \multirow{5}{6em}[0em]{\centering Provenance}
 & \cellcolor{black!10} Multi-resolution summaries 
\\
 &
 &
 & \cellcolor{black!10} Access patterns 
\\
 &
 &
 & \cellcolor{black!10} Observed performance
\\
 &
 &
 & \cellcolor{black!10} Cross-dataset correlations
\\
 &
 &
 & \cellcolor{black!10} Audit log of modifications
\\
\hline
\end{tabular}
\caption{Classification of metadata based on who provides/acquires
metadata and where metadata is stored. Human-generated metadata is acquired
using ad-hoc annotations stored either in `readme.txt' files, scientists'
logbooks, etc., and using high-level I/O libraries such as HDF5, NetCDF, and
ADIOS. Workflow systems acquire provenance related metadata automatically. Our
vision for rich metadata, such as summaries, data usage patterns, performance,
etc. is highlighted to acquire automatically.}
\label{tbl:richmeta}
\end{table*}

\subsection{A classification of metadata information}

From our interactions with scientists across different domains, we have
observed that there is substantial diversity in the metadata information
that scientists interact with. 
In Table~\ref{tbl:richmeta},
we show a classification of metadata types based on who provides metadata and where
metadata is stored. The common metadata
acquisition sources are scientists or application developers (labeled as
`Human-generated') and software libraries and tools (`Automatically acquired').
Metadata is often stored within the data files (labeled as `Integrated') or in separate
files or database systems (`Segregated'). 
We briefly discuss these categories in the following paragraphs. 

\textbf{User-defined metadata prior to data generation:}
Starting from the top right corner in Table~\ref{tbl:richmeta}, some
metadata are provided explicitly by a scientist at the moment of data
production.
Examples include data type and endianness information, or the
dimensionality and chunking strategy of an array for defining 
the layout of data on a disk.
Under this paradigm, a particular file format, such as HDF5, or data encoding defines
the collected metadata in advance and stores the metadata within the
dataset. For example, in HDF5 file format \cite{HDF5}, metadata is stored 
as the attributes of the file. 
Existing file format libraries already provide support to capture and
propagate this type of metadata; otherwise the dataset would be
indecipherable for analysis.

\textbf{Ad-hoc metadata that annotate a dataset:}
Another way to acquire metadata is to explicitly
request it by the domain expert. 
Many times, however, important information that describes specific
features or properties of the data cannot be shoehorned into this rigid
format. Scientists describe these properties in plain text that
accompanies the dataset.
As shown in the top left corner in Table~\ref{tbl:richmeta},
such metadata can be logically thought as an annotation to an existing
dataset which is stored separately from the data.
Common techniques that use this metadata management paradigm 
include publishing \texttt{readme.txt} files at the same website or
file system folder as the dataset, as well as electronic or face-to-face
communication. In certain instances this form of metadata is captured in
the logbooks of scientists and can only be acquired through direct communication.

\textbf{Workflow-based metadata:}
Several scientific workflow systems are in use to automate task
coordination and management in complex workflows.
Workflow systems collect information that is mainly
related to the provenance of the data and is used to quickly identify
(or repeat) the process that generated a specific fragment of a dataset.
Each workflow system stores and manages provenance metadata differently,
and no standardized interface to access lineage information from
applications currently exists.
Should a scientist desire to access the provenance metadata,
they first need to learn the data model and the query interface of the
workflow system.
Some workflow systems store provenance metadata in a relational
database and scientists can query the database using SQL \cite{MPO}. 
Other systems use representations that are
optimized for provenance metadata, and scientists query provenance
metadata through a system-specific API \cite{karma}.

\subsubsection*{Towards information-rich metadata}

In the extreme-scale data era that many scientific domains are entering,
it becomes necessary to automatically collect information-rich metadata
that go beyond provenance. 
Rich metadata include multi-resolution snapshots of data, summarizations, and
semantic relationships among datasets. 
We classify the new types of metadata that can be automatically 
captured in five broad categories:

\begin{enumerate}
\item
{\bf Identity information} 
includes the dataset name or any other unique identifier, the application
producing the dataset, and reproducibility information such as the
task and its parameters that were used to produce the result.

\item
{\bf Dataset descriptions}
include summary statistics,
derived variables, the resolution of the dataset, and the location of
low-resolution snapshots of data for quick visualizations. 

\item
{\bf Performance and profiling information}
are historical access patterns, the 
presence of augmented datasets (such as indexes, partial replicas, or materialized
views),
and the locations of physically reorganized
layouts of the data for faster access.
This category also includes prior response time and energy
consumption measurements that can be used for optimization decisions. 
This information can be readily leveraged for
exploratory scientific data analysis \cite{LifeRaft:Randal}.

\item
{\bf Relationships}
among various datasets or tasks, such as derived variables of a dataset,
or possible computations to derive such a variable upon request.
This includes information on different views of the same dataset, such
as a sorted replica or a bitmap index.
Relationship metadata 
captures how analysis results are computed and
where the results are stored. 

\item
{\bf User-defined metadata.} 
Users and applications will be able to specify additional metadata to
extend the semantic information conveyed as metadata. 

\end{enumerate}


\section{Science drivers}
\label{sec:app-drivers}

The proposed vision for integrated metadata management is useful for
a wide variety of scientific applications. In this section, we use
applications from plasma physics, climate modeling
and neuroscience to motivate the need and the potential benefits of
information-rich metadata. 

\subsection{Plasma Physics}
\label{sec:app-drivers:physics}

Scientific simulations in various fields of physics use large-scale
supercomputers for understanding the phenomena that affect our
universe.
Collisionless magnetic reconnection is one such important phenomenon
that releases energy explosively as magnetic field lines break and
reconnect in plasmas. This is a mechanism responsible for 
the aurora when the Earth's magnetosphere reacts to solar eruptions.
Magnetic reconnection is 
initiated in the small scale around individual electrons but eventually
leads to a large-scale reconfiguration of the magnetic field. Recent
simulations~\cite{Roytershteyn:2012} have revealed that electron kinetic
physics is not only important in triggering reconnection,
but also in its subsequent evolution. 

Fueled by new capabilities of highly-optimized simulations,
large-scale computers have been providing the
first glimpses of various physics phenomenon in a high resolution and
multidimensional space. 
The amount of data produced by these simulations is massive.
For instance, a recent execution of the VPIC simulation on
120,000 CPU cores of the Hopper supercomputing system at the National Energy
Research Scientific Computing Center (NERSC) produced 400 TB of
data. Yet, this massive amount of data corresponds to only 10 out of
the 23,000 simulated time steps! 
To counter the data volume problem, 
scientists often
perform various analyses while the data is in memory and discard
intermediate data. This analysis process is called \emph{in situ} and is
sometimes used for reducing the amount of data to be stored.
The complete particle dataset will be processed and
stored for only one or two final time steps instead. 
The burden of managing and orchestrating the \emph{in situ} analyses,
however, lies with the scientists that use existing processing
libraries, such as ADIOS \cite{adios} and Glean \cite{glean}. 
The absence of standardized metadata management
is an obstacle to automated coordination and synthesis of multiple
discrete \emph{in situ} analyses. 

Complicating the picture further, these \emph{in situ} analyses will
need to be continuously maintained and augmented to be conscious to
future changes in the memory and storage hierarchy.
Upcoming extreme-scale architectures 
will likely include technologies such as
large main memory, non-volatile memory, and flash-based storage in the
form of burst buffers for simulation and staging nodes. Making use of
these memory and storage resources is necessary for energy
efficiency and high performance. Storing metadata regarding optimizations
previously applied in improving these massive data I/O and results from previous
analysis will benefit not only performance, but also scientific productivity.

\subsection{Climate Modeling and Analysis}
\label{sec:app-drivers:climate}

Efficient collaboration among climate scientists is critical in
accelerating the understanding of the effects of climate change on
future generations. 
The number of scientists involved in climate modeling
and analysis is enormous. For example, there are more than 1,200
scientists in the IPCC's AR-5 Working Group II alone \cite{AR-5_number}.
Climate research has a long history of heavily relying on metadata to
interpret scientific observations.
The climate research community has embraced self-describing and
machine-independent data formats, and the netCDF data format \cite{netcdf:1990} in
particular is widely accepted for disseminating scientific datasets.

Multiple standardization efforts exist to encourage the exchange of
climate data observations and analyses.
The National Oceanic and Atmospheric Administration has proposed the
Cooperative Ocean/Atmosphere Research Data Service (COARDS)
\cite{COARDS} conventions to encourage the standardization of the
metadata for global atmospheric and oceanographic
research data sets. 
The Program For Climate Model Diagnosis and Intercomparison (PCMDI),
that manages the CMIP3 and CMIP5 model data has proposed the Climate and
Forecast (CF) Metadata Convention to generalize and extend the
COARDS conventions for datasets that use the netCDF API \cite{CF_metadata}.
In a similar effort, the Numerical Model Metadata (NMM) initiative
\cite{NMM} has standardized the description of the
numerical model used to produce climate data. This includes information
such as a formal and comprehensive description of the model and the
parameter settings used to run the model.
Other efforts have provided interfaces to annotate datasets with
particular types of metadata. These include
the Earth System Modeling Framework (ESMF) \cite{ESMF}, the PRogram for Integrated
earth System Modeling (PRISM) \cite{prism}, and the Earth System Grid (ESG)
\cite{ESG}. 
To encourage meta-analysis and synthesis studies, the 
Earth System Curator \cite{curator} proposed a metadata
infrastructure that combines and curates metadata from commonly-used
packages for climate modeling.

Despite these standardization efforts, extensive user involvement
is needed to add metadata and provenance information. 
In addition, the quality and type of metadata that are provided
depends entirely on the scientists and the conventions of the
organizations that provide the data. 
For example, in the CMIP5 datasets, the variance in the number of days
reported per year is diverse. Some models uniformly use 30 days for
every month, which amounts to 360 days per year.
Other models represent the number of days in each month accurately, 
but omit a leap day in February to simplify year-over-year analyses,
which results every year in a dataset having exactly 365 days.
In the meantime, raw observation data from sensors commonly
handle leap years correctly. 
While all these models use the same conventions to store the metadata,
climate data analysis scientists still have
to adjust and tune their analysis applications based on their
understanding of these conventions. 
Multiple other incompatibilities exist
that burden climate data scientists further. 
As a consequence,
many climate research datasets are annotated selectively after the analysis has
determined that the findings have sufficient scientific significance to
merit the additional effort. 

Similar to the VPIC magnetic reconnection application, climate simulations also can
take advantage of \emph{in situ} analysis. For instance, climate model
simulations write data to disk in periodic intervals. 
Extreme events, such as tornados, may quickly form and then disappear before
writing a snapshot of the simulation to disk. 
Interesting features between those intervals may be lost, and increasing
the frequency of the writes may be impractical.
These events can be detected dynamically through \emph{in situ} analyses.
Storing and managing metadata of these in situ detections for further inspection
will improve the quality of data significantly.

\subsection{Neuroscience}

Recent initiatives to better understand the human brain \cite{BlueBrain,
BrainInit} have spawned research efforts that collect enormous
amounts of neuroscience data.
The sources of
neuroimaging data span a wide variety of instruments and techniques
including computed tomography (CT), diffuse optical imaging (DOI),
magnetic resonance imaging (MRI), functional magnetic resonance imaging
(fMRI), Positron emission tomography (PET), 
etc. These instruments produce data in different formats depending on
their manufacturer. In addition, 
metadata corresponding to each data set examining the same subject or
sample are highly likely to be different as well. In this scenario,
extracting and managing relationships among different images of the same
subject/sample is a critical requirement to improve the understanding of
neuroscience data. 
We are not, however, aware of any proposal to automatically extract and manage
relationships among scientific datasets in a systematic manner. 

After the extraction of metadata and their relationships, the
requirement of searching them is
another challenge. Several solutions targeting bio-medical imaging
specifically have been proposed \cite{BIMM, BioDB}. Systems such as MPO
\cite{MPO}, the SPOT suite \cite{SPOT}, the JGI Archive and Metadata
Organizer \cite{JAMO}, and ESGF \cite{ESG} have been actively
providing metadata search capabilities. While the data remains in these
systems, searching data and metadata is possible. However, once the data
is outside these systems, searching metadata becomes
cumbersome. In fact, in 
many cases the metadata of interest may not be stored within the data file. 
To search the metadata, scientists have to develop custom solutions for
accessing these external metadata sources.
A requirement for a metadata management framework is to 
portably manage metadata and support 
movement, extraction, and search capabilities.


\section{Research Challenges} 
\label{sec:approach}

The implementation of an integrated metadata 
management framework 
to support various types of scientific analyses
faces several fundamental challenges
that span the acquisition, storage, and access of metadata information. 
In this section we describe some foreseeable challenges and
elaborate on possible solutions, in response to the following research
questions:

\begin{enumerate}
  \item What metadata has to be captured to make \emph{in situ}
and post-processing analyses more efficient? How does one collect
metadata non-intrusively from different sources?
  \item 
  How should one store the collected metadata and provenance for efficient
access? 
How can one control the storage space of information-rich
metadata to ensure that the size of the dataset remains manageable?
  \item How does one keep the metadata consistent and resilient in the
  presence of failures? How can one keep the metadata secure and control
every access?
\end{enumerate}


\subsection{Acquiring metadata}
\label{approach:acq}

\subsubsection{Integrating automatic metadata acquisition into existing
infrastructure}
Collecting a comprehensive set of metadata requires tapping into
multiple sources. The research challenge is identifying the appropriate
component of the infrastructure stack to acquire the metadata of interest.
Although it is easier to glean information-rich metadata from components
that users directly interact with, such as analysis or visualization
tools, this ecosystem is very diverse and segmented per
scientific discipline.
Low-level components of the stack, such as the file system or operating
system, are ubiquitous among large-scale computing facilities but
the acquired metadata may not have enough semantic information to be
useful.
We identify four abstract levels to intercept and acquire metadata for
scientific applications:

\begin{enumerate}
\item
{\bf Analysis level:} 
Observing metadata information at the application level allows the
framework to capture high-level information such as the analysis intent.
Because of the diversity of the modern scientific toolbox among
different scientific disciplines, acquiring high-level metadata requires
user involvement. Commonly, the domain expert will express concepts and
relationships using an established knowledge ontology, which may impede
the productivity of non-experts users of the data.
\item
{\bf Library level:} 
Scientific analyses are rarely building every component from scratch;
they instead rely on a common set of libraries to provide some core
functionality. 
Examples of such libraries include BLAS for linear algebra, MPI for
parallelization, and SQL connectors for issuing queries against a database.
A metadata management framework could integrate with these libraries
and transparently acquire metadata. The metadata information at this
level is commonly rich in semantics (for instance, a BLAS function call
translates to a single linear algebra operation), is high-level (for instance, a
SQL query specifies what data should be returned, and not how they will
be processed) and contains valuable details about the execution
environment (for instance, instrumenting the MPI library can provide a
wealth of metadata information on parallelization granularity and
communication patterns).
The challenge is that metadata information acquired at the library level
will be partial and fragmented because only a limited number of
libraries can be augmented with metadata acquisition capabilities.
\item
{\bf File format level:}
Looking further away from the user, one finds that data resides in
binary file formats following established specifications to maximize
sharing and reuse. File formats such as netCDF, HDF5, ADIOS
BP or SAM/BAM are \emph{de facto} standards for data exchange
among certain scientific communities.
As all I/O is issued against a well-defined API that is
specific to each file format library, a metadata management framework
could conceivably glean metadata information from access requests that
are directed towards the library. 
Some libraries, such as HDF5, are supporting this functionality
natively through a virtual object layer (VOL) abstraction \cite{VOL:2011}.
Metadata acquisition libraries that dynamically instrument codes can support
other libraries such as POSIX-IO.
An advantage of this technique is that metadata is acquired
automatically without recompiling existing programs.
The disadvantage is that data access requests at the file format level are
imperative and stripped of semantic meaning, as requests retrieve specific
columns of a table or cells in an array.
Another drawback is that certain file formats, such as FITS or CSV files
are not amenable to automated metadata acquisition as they lack an
established programming interface to access data. 
\item
{\bf Operating/file system level:}
A metadata framework can also tap into the mature APIs that are
available by the file system and the operating system. In addition to
tracing system calls (for instance, using \texttt{strace}), a framework
can also observe detailed hardware performance counters and process
statistics.
Identity metadata information can be gathered from the process
information and the computational infrastructure that is known to the
operating system. 
The appeal of such a mechanism is that all analysis tasks can be
observed without any code modification or involvement by the domain
expert. However, at this layer only physical actions on byte streams can
be observed.
Logically equivalent actions that change the encoding but not the
actual data (such as converting temperature from Celsius to Fahrenheit)
cannot be readily identified at this level.
\end{enumerate}

\subsubsection{Discovering relationships between datasets}
A significant research challenge is in capturing
cross-dataset relationships and correlations as metadata. 
Analysis tasks today require manual orchestration by domain experts to
fully exploit these relationships among data objects.
The first cross-dataset property of interest is whether two or more
datasets correspond to similar physical observations, yet are treated as
separate datasets because of nuances of the data production process.
An example of this type of cross-dataset relationship commonly arises in 
neuroscience, when analyzing a single brain specimen with two different
instruments results in two files in different proprietary formats that
scientists need to then analyze jointly and treat as a single dataset.
Another cross-dataset property of interest is derived variables that
require the concurrent analysis of different datasets.
For example, relating pressure with wind velocity in a climate dataset
for a particular geographical area may involve accessing data objects
with different resolutions, that are produced by teams following
different conventions and are stored in different file formats. 

Capturing relationships among different datasets can be a semi-automatic
approach. This may need some
involvement from scientists in defining the basic mapping of variables.
After the mappings are defined, during data production and analysis
phases, runtime libraries can be used to extract the metadata
information.


\subsection{Storing and accessing metadata}
\label{sec:meta-mgmt}

\subsubsection{Embedding metadata into datasets}

The cornerstone of our vision for scientific metadata management is that
metadata needs to be stored within datasets to prevent
metadata-oblivious processes from destroying metadata information.
This will accelerate the transition towards metadata-rich scientific
processing pipelines. 

Many scientific datasets are encoded in file formats that allow multiple
data objects to be stored within a file.
Embedding information-rich metadata objects in these file formats
will require devising new object hierarchies for metadata.
This requires metadata information to be discovered, encoded and
accessed using the established interfaces of the underlying file
format.

\begin{enumerate}
\item
{\bf Metadata discovery:} 
By embedding metadata within the data object hierarchy of a file format,
metadata discovery needs to build on and extend existing data object discovery
mechanisms.
Metadata information, however, that masquerades as a regular data object
may be accessed continuously during an analysis.
Metadata discovery includes querying about the existence of a particular
metadata class (``is there a summary of this dataset?'') and navigating
between metadata objects (``switch to a low-resolution summary for this
analysis'').
Popular file format libraries are implemented with the assumption that
data object discovery is a rare event, as it occurs only at the
beginning of the analysis.
In addition, object-to-object navigation commonly requires node-to-node
hops within tree-based hierarchies.
More efficient metadata discovery and navigation mechanisms need to be
designed for existing file format libraries to efficiently support
exascale metadata-rich applications. 

\item
{\bf Metadata encoding:}
Scientific file formats are commonly designed around array-centric data
models.
Some metadata types such as multi-resolution summaries are naturally
represented as arrays.
Metadata information needs to be encoded in array-like objects in order
to embed richer metadata within scientific file formats.
The open research challenge is to devise efficient representations for
different metadata classes, such as audit logs, that may lack a well-defined
structure.

\item
{\bf Metadata access:} 
Scientific metadata information that is stored within scientific file
formats is indistinguishable from regular data objects.
The advantage to applications is that they can access metadata using the
established data access interface.
This interface, however, needs to be augmented to expose the concurrency
that will be available at both the cluster level and the node level in
future computing infrastructure.
(For instance VPIC, the magnetic reconnection simulation described in
Section~\ref{sec:app-drivers:physics}, already uses 120,000 CPU cores
and is likely to approach the one-million-core scale in the future.)
Existing data access mechanisms rely on data structures and
communication patterns that are inherently non-scalable.
A promising research direction is to investigate and resolve the
scalability challenges that will allow metadata access interfaces to
sustain the rate and volume of metadata requests that will be triggered
by extreme-scale computing infrastructure. 

\end{enumerate}

\subsubsection{Transactional semantics for metadata}

Although file format libraries are designed to handle concurrent data
operations, the correctness semantics of concurrent modifications are
rarely described in the documentation.
This is in part due to the complexity of describing what correctness
means for all underlying components of the software stack, down to
the operating system and the file system.
This has not been a major concern for scientists so far because
large-scale analyses are data-parallel and rarely update existing data.
The append-only I/O pattern does not cause conflicts, and data
partitioning isolates any infrequent updates to existing data objects.
Unfortunately, metadata are shared by all tasks that operate on a
dataset and will be continuously updated with new information.

A metadata management framework for scientific data needs to understand and
expose the correctness semantics of concurrent metadata modifications to
a scientist.
Resilient systems rely on the notion of \emph{transactional
data operations} for this purpose. 
A transactional interface gives applications a consistent view of the
metadata, and allows metadata modifications in complete isolation. 
This significantly simplifies the application logic,
and allows the metadata management framework to perform sophisticated
and holistic optimizations to an entire batch of metadata operations for
efficiency.

Defining a transactional interface for metadata management leads to 
research questions about the foundations of transaction processing and
concurrency control \cite{GrayReuterBook, WeikumBook}.
The tuple-centric concurrency model of relational database systems is
not sufficient for information-rich metadata objects.
The first fundamental question is what is the unit of concurrency for
rich metadata information. The second fundamental question is what is
the set of permissible actions on the unit of concurrency, and what are
their associated actions under the do/undo/redo protocol that will
permit metadata operations to revert their effects.
This naturally leads to the question of what does it mean for two
transactional operations to conflict when updating metadata.
What is an acceptable conflict resolution depends on the type of the
metadata information and the scientific use case: conflicts in the
dataset description or identity information may be intolerable as they
can render a dataset inaccessible,
but update conflicts in the data access pattern profile may be
permissible if this information is only used for runtime optimization.

Controlling concurrent operations on metadata requires revisiting 
the fundamental techniques of \emph{single-version locking} and
\emph{multi-versioning} and extending them for
scientific metadata management. 
Synchronizing concurrent modifications via locking has high overhead,
but always retains and returns the single latest version of metadata
information.
By design, multi-version concurrency control (MVCC) does not destroy or alter
the original metadata when an application updates the metadata; the
edit is applied to a new copy of the data instead. 
Retaining multiple versions can improve performance by redirecting reads
and update requests to a different copy of the data
\cite{LarsonBDFPZ11}.
In addition, when scientists wish to inspect older versions of a dataset,
prior versions of metadata can be leveraged for answers.
(This feature has been referred to as ``time travel'' querying in prior
work \cite{SoroushB13}.)
Although reading a historic version does not violate the transactional
semantics, it is unclear if this is desirable by scientific applications
that act on the metadata information. 
Appropriate interfaces need to be developed to allow domain experts to 
specify their metadata freshness requirements for different types of
metadata and different scientific analyses.

\subsubsection{Storage size considerations}

A metadata management framework also needs to determine what
metadata should be discarded and when.
One trivial solution would be to tie the life cycle of metadata to the
life cycle of the associated data object: metadata collection starts when
a data object is first created, and metadata are discarded when an
object is deleted. However, there may be applications, such as security
auditing, where it is necessary to preserve metadata even after the data
has been discarded. The relative importance of different metadata
information is only known by the domain expert.
We therefore envision a scientific metadata management framework that
allows a domain expert to specify a user-defined threshold per
metadata class that is expressed as a fraction of the size of the base
dataset.
If the metadata size approaches the user-defined threshold, a metadata
management framework can explore a number of options to reduce the
storage footprint of the collected metadata.

\begin{enumerate}
\item
{\bf Selective disposal of metadata information:}
One alternative to reduce the storage size of metadata information to 
is to selectively dispose metadata information. However,
not all metadata information is equally valuable.
It is thus natural to decompose this task into two orthogonal
decisions: 
First, the framework needs to decide on the class of metadata that will
be the ``victim'' for metadata disposal. The domain scientist can protect a
class of metadata information through a \texttt{NEVER} directive, or
indicate a strong preference to truncate metadata information from a
particular class using an \texttt{ALWAYS} directive.
Second, the framework needs to decide the metadata information from the
``victim'' class that will be discarded. The historical metadata access
pattern could be used to identify infrequently used metadata and guide
the decision.

\item
{\bf Semantic compaction of metadata information:}

Another alternative is to semantically compact metadata to increase its
information density. 
One form of compaction is entry consolidation: 
for example, an audit log that tracks accesses to individual array cells
can cluster these accesses and express them as one access range.
Compaction can also be achieved by computing metadata information on the
fly from existing metadata. 
An example would be generating summaries at a desired resolution by
summarizing higher-resolution datasets or through linear interpolation
of lower-resolution datasets.

\item
{\bf Metadata-aware deduplication:}
Future exascale systems will process massive scientific datasets.
Because it will be very time-consuming to transfer datasets of this
volume across computing facilities, data will be predominantly analyzed
in the facility where they are stored.
In this scenario, it will not be uncommon to find multiple copies of a
dataset in a large-scale computing facility, often with minor
differences, in the project directories of different teams.
This exascale collaboration pattern presents an opportunity to leverage
dataset similarities for automatic metadata-aware deduplication in the
file system layer. These techniques extend the block-based approaches that are
automatically applied by storage subsystems today and differ in that
they will deduplicate at the logical (data) level, and not at the
physical (byte) level.
To effectively support logical deduplication, an exascale scientific
metadata management framework needs to determine dataset lineage
based on metadata.
The research challenge is to develop efficient metadata-aware
\emph{diff} and \emph{merge} algorithms that can quantify the lineage
confidence between different datasets based on their metadata
information. In addition, approximate algorithms must be developed that
rely on incomplete metadata information to accommodate different
metadata acquisition and metadata disposal practices.

\end{enumerate}

\subsubsection{Resilience and metadata consistency}
\label{sec:resilience}

As scientific applications like magnetic reconnection simulation
 approach the
million-core scale, a single node failure becomes nearly certain.
Although hardware failures remain relatively rare occurrences in
supercomputers today, they are
widely anticipated to become more common in future extreme-scale computing
infrastructure. 
Analyses will be also interrupted for more mundane reasons, such as
software bugs or transient communication failures. 
Workflow management systems can selectively repeat an analysis after
failure, but rely on metadata to decide what tasks need to be restarted.
Therefore, the consistency of the metadata information after a failure
remains of paramount importance.

A key design goal of our metadata management solution is to
tolerate and recover from hardware, software, or user errors graciously
and systematically. 
A na\"{\i}ve solution to achieve durability is to store all metadata on a
non-volatile storage medium, such as a hard disk, and write to this
medium on every modification. 
The unprecedented concurrency of emerging scientific workloads will trigger
millions of minuscule metadata modifications per second. 
Although the metadata information is safe in a case of a failure,
waiting for the data to reach stable storage on every update has three
major drawbacks: (1) it increases the response time for
every metadata modification, (2) it is energy inefficient, as it cannot
take advantage of caching, and (3) it drastically reduces the usable
lifetime of modern non-volatile storage media like SSDs that support a
limited number of write-erase cycles per cell.
A metadata management framework needs to adapt write-ahead logging for
the demands of extreme-scale applications, and evaluate
high-availability techniques that are based on replication in a
supercomputing environment.


\subsubsection{In-memory metadata caching}
\label{sec:storage}

A scientific metadata framework needs to provide capabilities to search
metadata efficiently. Existing systems such as ESGF \cite{ESG} and JAMO \cite{JAMO}
provide searching metadata based on text, where the amount of metadata is 
minimal. Searching for extensive and
information-rich metadata 
such as data summaries or performance metadata, requires
an efficient storage organization of the information. 
Passively storing and retrieving metadata from disk-based storage, which
has been the current practice for metadata management in workflow
systems, 
will be prohibitively slow for emerging scientific use cases such as
interactive scientific data visualization and real-time data analytics. 
Poor performance will be exacerbated by the
unpredictable remote disk access latency in large-scale computing
environments. 
Future metadata management frameworks will need to optimize metadata
storage for low access latency. It is possible to achieve this goal
using transparent architecture-conscious metadata placement in the
memory and disk storage hierarchy. 

The increasing degree of parallelism of future supercomputers will require
novel storage techniques for metadata management.
One can reasonably expect that applications will reserve the majority of
the memory for scientific data analysis or simulation, and only a sliver
of memory will be available for other information, such as metadata.
Given the severe memory constraints, a promising research direction is
to study data placement mechanisms and policies for metadata.
We bring attention to three aspects of this problem: 

\begin{enumerate}
\item
Given an analysis or simulation task and its configuration parameters,
what is the most relevant metadata to speed up the scientific discovery
process?
\item
When should the relevant metadata be retrieved from non-volatile
storage and on what node should they be placed?
\item
When should metadata be evicted from memory, and where should they
be stored in the non-volatile storage hierarchy?
\end{enumerate}

A light-weight prediction model can be used to
identify the relevance of each metadata object
for processing, depending on the type of the task and
the data fragment it is processing.
The model can 
leverage prior statistical knowledge of metadata accesses for 
smart metadata placement and prefetching.
Novel hardware primitives such as hardware transactional memory, can
further simplify application logic and improve
performance \cite{TranBN10}.

\subsubsection{Metadata confidentiality and access control}

Even if a scientific dataset is publicly available, the metadata pertaining to this
dataset must be carefully disseminated by the metadata
management framework.
Metadata information that relates to the production of the data and its
processing would inadvertently disclose information about the data
production process, which may be confidential.
Relationship metadata and multi-resolution summaries of a public
dataset may reveal the existence of related but confidential datasets,
while stored performance profiles may disclose confidential
infrastructural capabilities.
If scientific metadata management is to become systematic and ubiquitous, a
framework must guarantee \emph{confidentiality} for metadata at rest, and
\emph{access control} to metadata in motion.

One research challenge is to develop a security model for metadata
at rest. This model would rigorously define metadata information
leakage and identify specific risk vectors. Based on this model, the
user will specify non-permissible metadata information that will be
discarded during acquisition based on the sensitivity of the analysis or
the dataset. 
Assymetric cryptography of the metadata at rest can 
prevent metadata leakage if media theft occurs.
In addition, the scientific metadata management framework needs to provide
tools for selectively scrubbing sensitive metadata from datasets during
dataset curation.

Additional research challenges arise when considering techniques that
maintain the confidentiality of metadata in motion but permit metadata
sharing between different analyses. An open research question is how
access control on metadata can be implemented effectively using existing
operating system and file system primitives, such as access control
lists or capabilities. One challenge is that the granularity of
control is commonly a file, yet large-scale scientific simulations
commonly store hundreds of datasets in a single petabyte-sized file.
A promising research avenue is devising fine-grained control mechanisms
for metadata management for exascale scientific applications.


\section{Related Work}
\label{sec:relwork}

The importance of metadata and provenance has been recognized in many
different domains.
In the Web domain,
the Semantic Web is a collaborative standardization effort to formally
represent knowledge. 
The Semantic Web community has created a
number of standards and software tools, such as RDF \cite{rdf}, RIF
\cite{rif}, OWL \cite{owl}, SPARQL \cite{sparql}, etc. to achieve this.
Despite these efforts, the Semantic Web has not yet been universally
embraced. 
Many efforts have proposed and developed for organizing scientific
metadata into domain-specific vocabularies
(``ontologies'') allowing machines to understand and extract the
semantic meaning of data \cite{OntologiesForSDM}. 
These efforts, however, have been limited to specific application
domains and standardization does not extend to generic scientific metadata.

Karma \cite{karma} is a stand-alone tool that collects and represents
provenance information for digital scientific data. 
The Karma tool can collect file-based provenance information by
automatically modifying batch job scripts 
\cite{karma_collection},
extract provenance information from log files 
\cite{karma_log},
and visualize provenance 
\cite{karma_visualization}.
Prior work has also suggested interfaces to access provenance
information. Karvounarakis et al. \cite{ProQL:2010} proposed ProQL, a
querying language for searching the lineage of data. Mani et al. 
\cite{Mani:2011} proposed query language constructs and Holland et al.
\cite{PQL:2008} developed an interface called PQL for querying
provenance. 

There have also been efforts to systematically manage metadata for
specific scientific domains. In the atmospheric sciences domain, 
Pallickara et al.~\cite{PallickaraPZ12} have proposed the Atmospheric
Data Discovery System (ADDS) to index and search observational datasets.
In the environmental research domain, 
the ```Data Near Here'' system 
allows scientists to reuse
and exchange data without relying on ``experts'' to convey 
what the dataset structure is \cite{MaierMBJST12, MeglerM11}.

Scientific metadata storage must be highly available and
resilient to failures.
Because metadata includes information on the physical location and layout
of scientific data, it may be impossible to understand the data if the
metadata is not accessible.
Many scientific data management
systems therefore store metadata in a relational DBMS for durability. 
SciDB \cite{SciDB} 
uses PostgreSQL for storing the
shape and the location of each dataset fragment.
The ``Data Near Here'' system also uses PostgreSQL for metadata storage
\cite{MeglerM11}.
ROARS, a data repository used by a biometrics research lab at
the University of Notre Dame \cite{Bui12}, stores metadata using
MySQL. 
Not all metadata, however, can be universally represented using a fixed
and structured relational schema. Hence, traditional relational database
systems can sometimes be unnatural choices for storing metadata that
cannot be efficiently represented using the relational data model.
The SPOT suite \cite{SPOT} uses MongoDB for storing light-source data and
simulations. The JGI Archive and Metadata Organizer (JAMO) \cite{JAMO} 
archives data related to gene sequences and also stores metadata in MongoDB. 

Prior work has studied efficient techniques to store and query generic
metadata information within a workflow. 
Shankar et al.~propose techniques to integrate scientific workflow
management in a relational database system \cite{ShankarKDN05}.
Bowers and Lud\"{a}scher have set the theoretic foundations for
propagating semantic annotations within \cite{BowersL06}.
Callahan et al.~have proposed VisTrails, a workflow system that supports
visual data exploration and visualization of provenance information
\cite{VisTrails}.
Bao et al.~have developed a dynamic labeling scheme that allows data
items to be efficiently labeled with fine-grained provenance
information \cite{BaoDM12}. Queries about whether two data items in
different phases of a workflow are related can be expressed as
reachability queries.
Finally, Heinis and Alonso have proposed to represent provenance dependency
graphs as intervals, and show how to query these intervals efficiently
\cite{HeinisA08}.


\section{Conclusions}
\label{sec:conclusions}

\emph{Ad hoc} scientific data exploration and analysis today relies on a domain
expert who can understand the structure of the data and the nuances of the
process that generated it.  
Integrated metadata acquisition and management can significantly
increase the reuse value of scientific datasets and the productivity of scientists. 
The efficiency of scientific exploration increases as data analysis
tasks can be directed more judiciously using the
insights stored and conveyed as metadata. 
This will accelerate
exploratory analysis of scientific data, and increase the longevity and reuse of curated
data and insights. 
Our vision for integrated and ubiquitous metadata management is a critical
stepping stone towards achieving physical and logical data independence for
extreme-scale scientific applications. 
\section*{Acknowledgments}

This work is supported in part by the Director, Office of Laboratory
Policy and Infrastructure Management of the U.S. Department of Energy
under contract DE-AC02-05CH11231, the U.S. National Science
Foundation under grant IIS-1422977 and a Google Research Award.

\bibliographystyle{abbrv}
\bibliography{paper}

\end{document}